\documentclass[fleqn,12pt,twoside]{article}
\usepackage{amsmath}
\usepackage{espcrc1}

\usepackage{graphicx}

\newcommand{\cx}{\rm{x}}
\newcommand{\cy}{\rm{y}}
\newcommand{\cz}{\rm{z}}
\newcommand{\ct}{\rm{t}}

\title{Collective dynamics of the fireball}

\author{Boris Tom\'a\v sik\thanks{Boris.Tomasik@cern.ch}\\[2ex]
{The Niels Bohr Institute, Blegdamsvej 17, 2100 Copenhagen \O, Denmark}}

\begin{document}

\maketitle

\begin{abstract}
I analyse the identified single-particle $p_t$ spectra and two-pion 
Bose-Einstein correlations from RHIC. They 
indicate a massive transverse expansion and rather short
lifetime of the system. The quantitative analysis in framework of the 
blast-wave model yields, however, unphysical results and suggests that 
the model may not be applicable in description of two-particle correlations.
I then discuss generalisations of the 
blast-wave model to non-central collisions
and the question how spatial asymmetry can be disentangled from flow asymmetry
in measurements of $v_2$ and azimuthally sensitive HBT radii.
\end{abstract}

\setcounter{footnote}{0}


\section{INTRODUCTION}

When studying nuclear collisions at highest energies we are 
interested in properties of strongly interacting matter. 
A clear signal of existence of an extended piece of 
matter is its collective expansion
which is not present in simple nucleon-nucleon collisions. This can
be deduced from the slopes of identified hadronic single-particle 
$p_t$ spectra:
while the slope is universal for all particle species in 
proton-proton collisions, the spectra become flatter with increasing 
particle mass in collisions of heavy ions.

Hadrons interact strongly an so they can only decouple from the system 
when it becomes dilute enough. Their spectra are fixed at the
moment of freeze-out and carry  information about the 
phase-space distribution in the final state of the fireball. 
We want to reconstruct this information and get an idea about 
the collective evolution of the fireball by comparing its final state 
to its initial state which is known from the energy and the 
collision geometry.

The process of freeze-out is very complicated \cite{magas}. Here I will 
simplify it and assume that all particles 
decouple along a specified three-dimensional freeze-out hypersurface 
\cite{cf}. The phase-space distribution at the freeze-out hypersurface 
will be parametrised by the so-called {\em blast-wave model}.

I will first introduce the model and analyse single-particle $p_t$ spectra
and two-pion correlations from RHIC. Then I show how the model is 
generalised to non-central collisions and focus on calculation
of $v_2$ and azimuthal dependence of HBT radii.


\section{THE BLAST-WAVE MODEL FOR CENTRAL COLLISIONS}

I will assume that in the end of its evolution the fireball 
is in a state of local thermal equilibrium given by temperature 
$T$ and chemical potentials $\mu_i$  for every species $i$.
The freeze-out hypersurface   will stretch along 
the $\tau = \sqrt{\ct^2 - \cz^2} = \tau_0 = \mbox{const}$ 
hyperbola\footnote{As space-time coordinates I will be using 
longitudinal proper time $\tau = \sqrt{\ct^2-\cz^2}$ 
($\cz$ is the Cartesian coordinate in beam direction) and space-time 
rapidity $\eta = 1/2 \ln((\ct+\cz)/(\ct-\cz))$. 
In the plane transverse to the beam, radial coordinates $r$ and $\psi$ are
used.} 
and will not depend on the transverse radial coordinate $r$. 
There is longitudinally boost-invariant expansion, i.e.\ $v_z = z/t$. 
The velocity field
\begin{equation}
u^\mu = (\cosh\rho\, \cosh\eta,\, \cos\psi\, \sinh\rho,\, \sin\psi\, \sinh\rho,
\, \cosh\rho\, \sinh\eta)\, .
\end{equation}
Here, $\rho$ stands for the {\em transverse rapidity} which will be assumed
to depend linearly on $r$
\begin{equation}
\rho = \rho_0 {r}/{R}\, .
\label{fpc}
\end{equation}
Note that this linear dependence was found for small transverse 
velocities also in hydrodynamic simulations \cite{Kolb:2003gq}. In the last
equation, $R$ is the transverse size of the fireball. The transverse density
distribution is uniform for $r<R$. 
The (Wigner) phase-space distribution of particles 
at freeze-out is then characterised by 
the emission function\footnote{Momentum is parametrised with the help of 
rapidity $y$, transverse momentum $p_t$, transverse mass 
$m_t = \sqrt{m^2+p_t^2}$, and the azimuthal angle $\phi$  such that 
$p^\mu=(m_t \cosh y,\, p_t \cos\phi,\, p_t\sin\phi,\, m_t\sinh y)$.}
\begin{eqnarray}
S(x,p) \, d^4x & = & \frac{m_t\, \cosh(y-\eta)}{(2\pi)^3}\, 
\left [\exp\left ( \frac{p\cdot u - \mu}{T}\right ) \pm 1 \right ]^{-1}
\theta(R-r)\nonumber \\
&& r\, dr\, d\psi\, d\eta\, \frac{\tau \, d\tau}{\sqrt{2\pi\, \Delta\tau^2}}
\exp\left ( - \frac{(\tau-\tau_0)^2}{2\, \Delta\tau^2}\right )\, ,
\label{ef}
\end{eqnarray}
where the term $m_t\, \cosh(y-\eta)$ comes from the Cooper-Frye pre-factor
$d\sigma^\mu p_\mu$ \cite{cf} and the exponential in $\tau$ represents
smearing in the freeze-out time. The energy in the statistical 
distribution is taken in the rest-frame of
the fluid: $E=p\cdot u$. This introduces  ``coupling'' between  momentum 
of the emitted particle and the flow velocity of the piece of fireball where
it comes from. The strength of this coupling is controlled by $1/T$.


\subsection{Calculation of spectra}

The single-particle spectrum is obtained from the correlation function 
as
\begin{equation}
E\frac{d^3N}{dp^3} = \int d^4x\, S(x,p)\, .
\end{equation}
For low $p_t$, the inverse 
slope\footnote{The inverse slope $T_*$ appears as a parameter of a 
fit to the $m_t$ spectrum with the function ${\cal N} \exp(-m_t/T_*)$.}
of the spectrum is roughly given by \cite{cl}
\begin{equation}
T_* = T + m \langle v_t \rangle^2\, .
\label{esl}
\end{equation}
Here, $\langle v_t^2 \rangle$ is the mean 
transverse collective expansion velocity. Thus the freeze-out temperature
and the transverse expansion cannot be both determined from a measurement 
of only one single-particle spectrum. However, they can be determined from 
identified spectra of different species.

The full expression for the single-particle spectrum reads \cite{ssh}
\begin{eqnarray}
E\frac{d^3N}{dp^3} & = & \frac{1}{2\pi^2} \int_0^\infty dr\,  r\, 
G(r)\,\tau_0(r)\, 
\Bigg \{ m_t \, K_1\left ( 
\frac{m_t\cosh(\rho(r))}{T}\right )
I_0\left ( \frac{p_t\sinh(\rho(r))}{T}\right ) 
\nonumber \\ &&
\qquad \qquad \qquad \qquad {}
- p_t \frac{d\tau_0}{dr}\, 
K_0\left ( \frac{m_t\cosh(\rho(r))}{T}\right )
I_1\left ( \frac{p_t\sinh(\rho(r))}{T}\right )
\Bigg \}\, .
\label{specfor}
\end{eqnarray}
Note that in the model given by emission function eq.~\eqref{ef},
$G(r) =\theta(R-r)$ and  the second 
term vanishes because $d\tau_0/dr=0$. 
Here I display a more general expression because I want 
to discuss all effects which influence the single-particle spectrum:
\begin{enumerate}
\item
Freeze-out temperature and the transverse expansion velocity. 
\item\label{p2}
Contributions to particle production from resonance decays. This is 
not included in the formula \eqref{specfor} but is important in particular
for pion production (and will be included in calculations below). In this 
connection also chemical potentials of the individual resonance species are
important as they determine the number of resonances and thus the 
fraction of pions which stem from their decays.
\item\label{hs}
The shape of freeze-out hypersurface enters through the term $d\tau_0/dr$.
\item\label{p4}
Also ``finer details'' like the profile $G(r)$, 
spatial dependence of the temperature, the radial 
profile of the transverse velocity, etc.\ can influence the spectrum
\cite{cl,fintun}.
\end{enumerate}
It is important to realise that two models which differ in points 
\ref{p2}--\ref{p4} may not give the same $T$ and $\langle v_t\rangle$ when 
fitted to single-particle spectra. For example, in 
the single-freeze-out model \cite{sfo}  
$\tau_0 = \sqrt{\mbox{const} + \cz^2 + r^2}$ and the second term on the 
r.h.s.\ of eq.~\eqref{specfor} gives a contribution. 
Thus the temperature
obtained from  fits with that model cannot be directly compared to 
the one which will obtained below \cite{uhbnl}.


\subsection{Calculation of HBT radii}

I will also calculate the HBT\footnote{The acronym HBT comes from 
names of the scientists who applied the method first in radioastronomy
in the 1950s: R.~Hanbury Brown and R.Q.~Twiss.}
radii in Bertsch-Pratt parametrisation 
(see e.g.~\cite{rev} for more details on HBT interferometry). They appear 
as width parameters of a Gaussian parametrisation of the correlation 
function
\begin{equation}
C(q,K) - 1 = \lambda\exp 
\left ( -R_o^2(K) q_o^2 - R_s^2(K) q_s^2 - R_l^2(K)q_l^2
\right ) \, ,
\label{gpar}
\end{equation}
where $q$ and $K$ are momentum difference and the average momentum of 
the pair, respectively, and the $R$'s are the HBT radii. They measure 
the {\em lengths of homogeneity} \cite{ms}, 
i.e., sizes
of the part of the source which produces pions with specified momentum.
The sizes are measured in three directions: {\em longitudinal} is 
parallel to the beam, {\em outward} parallel to transverse component 
of $K$, and {\em sideward} is the remaining Cartesian direction.
The parametrisation~\eqref{gpar} is valid in the CMS frame at mid-rapidity 
of central collisions; otherwise terms mixing two components of
$q$ may appear (like $R_{ol}^2 q_o q_l$, for example; see \cite{rev} for
more details). The HBT radii will be determined as 
\begin{subequations}
\label{modind}
\begin{eqnarray}
R_o^2(K) & = & \langle (\tilde \cx - \beta_t \tilde\ct)^2\rangle 
\, , \\
R_s^2(K) & = & \langle \tilde\cy^2 \rangle\, , \\
R_l^2(K) & = & \langle \tilde \cz^2 \rangle \, .
\end{eqnarray}
\end{subequations}
In eqs.~\eqref{modind}
$\cx$, $\cy$, $\cz$, $\ct$ stand for the Cartesian space-time coordinates,
$\cz$ is the longitudinal coordinate and $\cx$ the outward coordinate.
Averaging and the tilde are defined as
\begin{equation}
\langle f(x) \rangle (K) = 
\frac{\int d^4x\, f(x)\, S(x,K)}{\int d^4x\, S(x,K)}\, ,
\qquad
\tilde x^\mu = x^\mu - \langle x^\mu \rangle\, ,
\qquad
\beta_t = \frac{K_t}{K^0}\, .
\end{equation}

Because the lengths of homogeneity depend on  momentum, the 
HBT radii show dependence on $K_t$. For $R_s^2(K_t)$, an 
approximate formula\footnote{Actually, this formula was 
derived in a slightly different model, where $R_G$ was the transverse 
size. Nevertheless, it can be used for the qualitative arguments 
here.} 
says \cite{cl}
\begin{equation}
R_s^2 = \frac{R_G^2}{1 + \rho_0 M_t/T}\, \qquad \mbox{with}
\quad M_t = \sqrt{m^2 + K_t^2}\, .
\end{equation}
Thus the temperature $T$ and transverse expansion measured by $\rho_0$
are coupled together here. However, recalling eq.~\eqref{esl}, they 
can be disentangled by analysing both the HBT radii and the single 
particle $m_t$ spectrum.


\section{THE FREEZE-OUT STATE IN CENTRAL COLLISIONS}

I will fit the model to single-particle spectra and HBT radii measured 
at RHIC. First, each of the identified spectra of pions, kaons, and 
protons of both charges will be fitted individually. This provides
a consistency check for the assumption that all these species freeze-out
simultaneously: if the results of the fits are incompatible, 
this assumption is invalid. After that, I will fit the
HBT radii and look again whether they can be accommodated with the 
same model as the single-particle spectra.

Resonance production of pions will be taken into account for spectra
but not for HBT radii. A study of the influence of resonance production
on HBT radii indicated that in the presence of transverse flow 
they are changed only marginally \cite{Wiedemann:1996ig}.

Chemical potentials of the resonance species are determined 
in accord with the model of {\em partially chemically frozen gas} 
\cite{Bebie:1991ij}. It assumes that after the chemical 
freeze-out \cite{Braun-Munzinger:2001ip}
the {\em effective} numbers of particles decaying weakly (and thus slowly) 
stay fixed while the strong (and therefore fast) interactions stay 
in equilibrium. Effective number of any given species includes this species
plus particles which can be produced on average from 
decays of all present resonances. Adjectives ``fast'' and ``slow'' refer 
to comparison with the typical time scale of the fireball evolution.
As an example: effective number of pions 
$\tilde N_\pi = N_\pi+ 2N_\rho + N_\Delta + \dots$, 
and chemical potential of $\Delta$
will be $\mu_\Delta = \mu_\pi + \mu_N$ because it is in equilibrium 
with pions and nucleons. 


{\bf Au+Au at $\sqrt{s}=130\, A\mbox{GeV}$.}
When fitting the various single-particle spectra \cite{ps130}, 
no overlap of the 
fit results is found at 1$\sigma$ level. The $\chi^2$ contours 
in freeze-out temperature $T$ and the average transverse expansion velocity 
$\langle v_t \rangle$ are overlapping at 95\% confidence level 
at $T\approx 120 - 140 \, \mbox{MeV}$ and 
$\langle v_t \rangle \approx 0.45 - 0.5$.

\begin{figure}[t]
\centerline{\includegraphics[width=14cm]{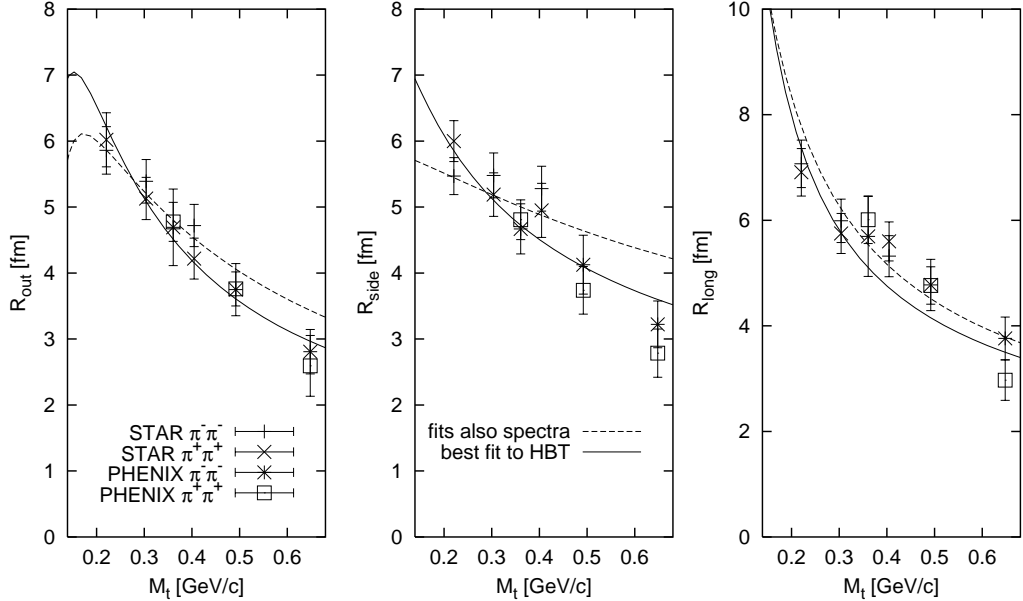}}
\vspace*{-1.5cm}
\caption{Two fits to the HBT radii measured in Au+Au collisions at 
$\sqrt{s} = 130 A\mbox{GeV}$. Dashed lines show a fit with values that 
also reproduce the single-particle spectra: $T=121\, \mbox{MeV}$, 
$\langle v_t\rangle = 0.51$, $R= 12.1\, \mbox{fm}$, 
$\tau_0 = 4.3\, \mbox{fm}/c$, $\Delta\tau = 4.4\, \mbox{fm}/c$. 
Solid lines show the best fit to $\pi^+\pi^+$ correlations:
$T=33\, \mbox{MeV}$, $\langle v_t\rangle = 0.73$, $R=24.1\,\mbox{fm}$,
$\tau_0 = 21.3\, \mbox{fm}/c$, $\Delta\tau = 1.1\, \mbox{fm}/c$.
\label{hbt130}}
\end{figure}
Can we fit the HBT radii with these values? In Figure~\ref{hbt130}
I plot the HBT radii measured by STAR \cite{starhbt130} and PHENIX 
\cite{phenhbt130} collaborations together with two fits. It turns out 
that $R_s(K_t)$ is so steep that---within this model---it rules out 
the temperatures obtained from spectra analysis. A lower
temperature is favoured, the best fit being obtained for 
$T=33\, \mbox{MeV}$! But this is not the freeze-out temperature!
The unphysically low value of the parameter $T$ just indicates that
in order to reproduce the observed strong $K_t$ dependence of HBT radii
the coupling of momentum to the local flow velocity in the 
emission function eq.~\eqref{ef} must be very strong.

\begin{figure}[t]
\centerline{\includegraphics[width=14cm]{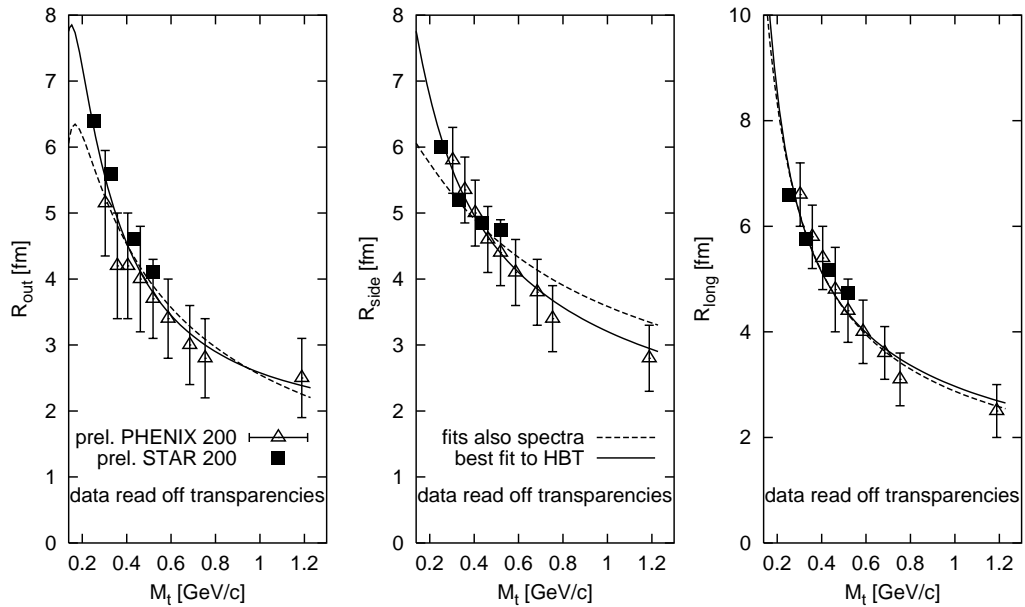}}
\vspace*{-1.5cm}
\caption{Fits to the HBT radii measured in Au+Au collisions at 
$\sqrt{s} = 200 A\mbox{GeV}$. Preliminary data are read off the
figures presented at the Quark Matter 2002 conference \cite{stqm02,phqm02}.
Dashed lines show a fit with values that 
also reproduce the single-particle spectra: 
$T = 115\, \mbox{MeV}$, $\langle v_t \rangle = 0.57$, $R=13.3\, \mbox{fm}$,
$\tau_0 = 4.7\, \mbox{fm}/c$, $\Delta\tau = 4.7\, \mbox{fm}/c$.
Solid lines show the best fit to the HBT radii: 
$T = 28\, \mbox{MeV}$, $\langle v_t \rangle = 0.71$, $R=27.7\, \mbox{fm}$,
$\tau_0 = 24.9\, \mbox{fm}/c$, $\Delta\tau = 1.7\, \mbox{fm}/c$.
\label{hbt200}}
\end{figure}
%
{\bf Au+Au at $\sqrt{s}=200\, A\mbox{GeV}$.}
Results from fits to identified single-particle spectra of 
pions, kaons and protons \cite{stsp200,phsp200} overlap at 
1$\sigma$ level at $T\approx 115\, \mbox{MeV}$ and 
$\langle v_t \rangle \approx 0.57$. The fit to HBT radii (Fig.~\ref{hbt200})
with these parameters is marginally good.
The best fit is again obtained with an unphysically low freeze-out 
temperature. 
The distinguishing power between the two models will be improved when 
final data will be fitted. Note that a fit to final results from 
PHENIX seem to be in better agreement with the conventional 
parameter values \cite{fabriceQM04}.

A very low freeze-out temperature is also obtained when the complete 
collection of HBT radii form Pb+Pb collisions at projectile energy 
of 158~$A$GeV is fitted. That fit is not shown here due to lack of space.


{\bf Conclusions from central collisions.}
The parameters obtained in the fits to  $K_t$ dependence of
HBT radii have no direct physics interpretation. The low temperature just 
indicates the need for very strong coupling between momentum and flow
within the blast-wave model. The model seems not to describe 
the realised scenario of freeze-out.

In spite of all these failures, one has to notice that the measured 
$R_s$, which is the rms radius of the source, indicates a strong transverse 
expansion since the original rms radius of Au nucleus is about 
3.5 fm. If we model the nucleus and the fireball by a box-like transverse
distribution, this is an expansion from a radius of $~7\, \mbox{fm}$
to about 12~fm (in the ``conventional'' models with $T\approx 100 - 120
\, \mbox{MeV}$). The total lifetime in Bjorken longitudinal expansion
scenario
can be estimated from $\tau_0 + \Delta\tau \approx 9\, \mbox{fm}/c$. 
It remains puzzling how the fireball expands transversely so much 
in such a short time without any original transverse flow.


\section{NON-CENTRAL COLLISIONS}
\label{noncen}

Next I generalise the model for non-central collisions. The transverse
profile becomes ellipsoidal instead of circular, so this changes
the $\theta$-function from eq.~\eqref{ef} into
\begin{equation}
\theta(R-r) \to 
\theta\left (R - R\sqrt{ \frac{\cx^2}{R_x^2} + \frac{\cy^2}{R_y^2}}\right )
\, , \qquad R_x = a\, R \, ,\qquad R_y = R/a\, .
\end{equation}
Here I define the direction of the $\cx$-coordinate to be parallel 
to the impact parameter, and $\cy$-coordinate 
as perpendicular to the reaction plane. 
The spatial anisotropy is parametrised by the parameter $a$. The transverse
flow rapidity---given by eq.~\eqref{fpc} for central collisions---will 
also vary with the angle $\psi$ with respect to the reaction plane
\begin{equation}
\rho(r,\psi) = \frac{r}{R} \sqrt{a^{-2} \cos^2\psi + a^2 \sin^2\psi}
\, \rho_0 \, \left (1 + \rho_2 \cos(2\psi_{\textrm{flow}})\right )\, .
\end{equation}
I will discuss two models which differ in the azimuthal dependence of 
$\rho(r,\psi)$ \cite{prpaper}. \\
{\bf Model 1:} Transverse velocity is perpendicular to the surface
and $\psi_{\textrm{flow}} = \mbox{Arctan}\frac{\cy}{\cx}$ \cite{fabrice}.\\
{\bf Model 2:} Transverse velocity points outwards in radial 
direction, $\psi_{\textrm{flow}} = \psi$.


\subsection{Elliptic flow coefficient $v_2$}

Elliptic flow can be generated by two different mechanisms in our model.
Firstly, the fireball can be spatially ellipsoidal, but have the same 
transverse velocity at the outer shell. Secondly, the fireball can have
spatially circular cross-section, but the expansion velocity can be
higher in one direction than in the other. It would be interesting to 
disentangle these two effects from the data. For example, if we can 
learn about the shape of the fireball, we obtain an indirect 
information about the lifetime because the fireball  
starts as out-of-plane elongated
and via expansion this shape becomes more and more circular. 
An out-of-plane elongated (i.e.\ $a<1$) fireball at freeze-out is
thus an indication of short lifetime.

The coefficient $v_2$ can be calculated in both used models; it turns out
that the results obtained for Model 1 map exactly to those for 
Model 2 if one replaces $a\to 1/a$ \cite{prpaper}. 
Hence, without  the knowledge 
about which model applies better for the description of the freeze-out,
one cannot say anything about whether the fireball is in-plane or 
out-of-plane elongated at freeze-out just by looking at $v_2$. 
If the model is fixed, $v_2$ depends on $a$ and $\rho_2$ in a different
way for different species, so one should be able to disentangle 
them from a measurement of $v_2(p_t)$ of more than one identified 
species.


\subsection{Azimuthally sensitive HBT}

Since the HBT radii are measured along directions specified by the
orientation of the transverse momentum, they depend {\em explicitly}
on the angle $\phi$ between the transverse momentum and the reaction plane. 
In addition to that, homogeneity regions depend on the momentum of 
produced particles and this introduces an {\em implicit} azimuthal 
dependence. (See \cite{rev} for more on azimuthal dependence 
of HBT radii.)

\begin{figure}[t]
\begin{minipage}{10cm}
\hspace*{-2cm}
\includegraphics[width=12cm]{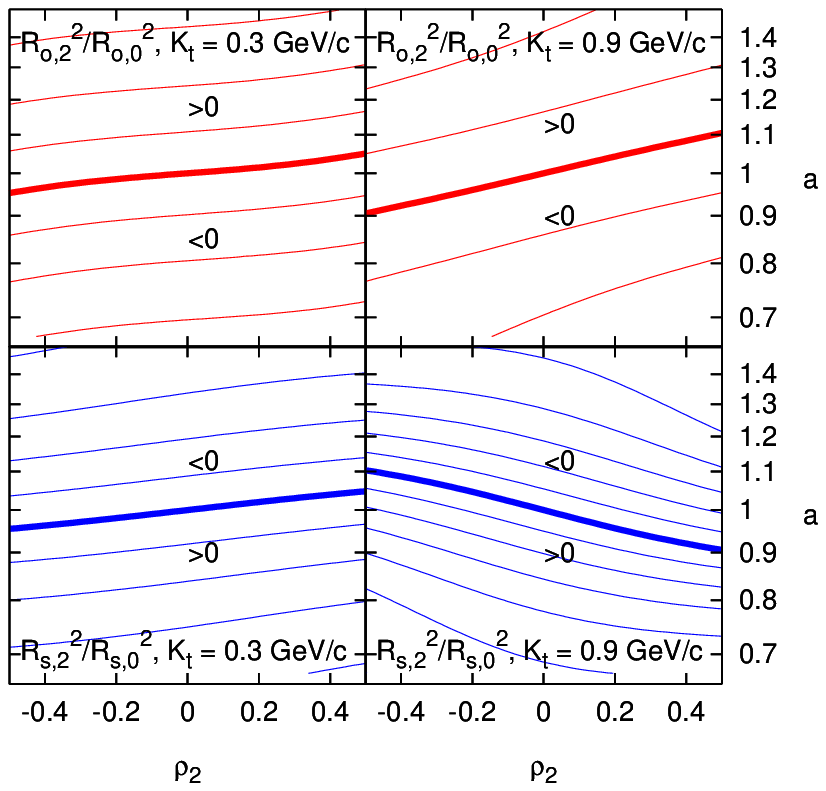}
\vspace*{-2.5cm}
\end{minipage}
\begin{minipage}{5cm}
\hspace*{-2cm}
\includegraphics[width=7.9cm]{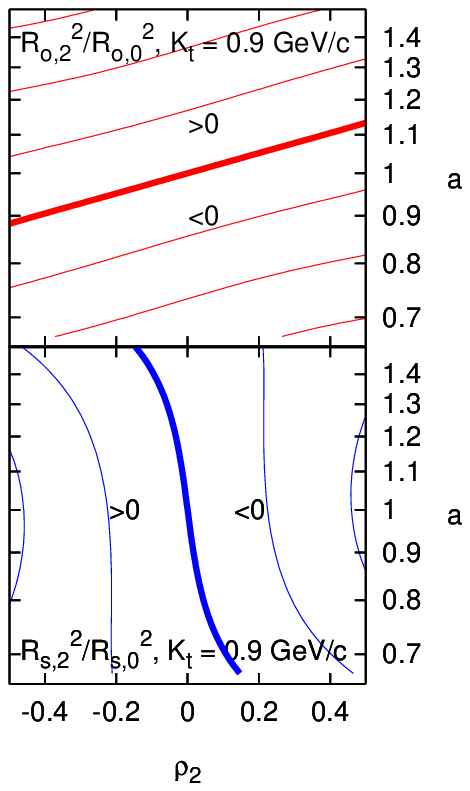}
\vspace*{-2.5cm}
\end{minipage}
\caption{%
The second order Fourier terms normalised by the average HBT radii
$R_{i,2}^2/R_{i,0}^2$ (see eq.~\eqref{r2}) as a function of $a$ 
and $\rho_2$. Upper row: outward radius; lower row: sideward radius.
Thick lines show where there is no second-order oscillation of the 
radius as a function of $\phi$. Consecutive lines correspond to 
increments by 0.1. Left panel: calculation with Model 1 at 
$p_t = 300\, \mbox{MeV}/c$ (left) and $p_t = 900\, \mbox{MeV}/c$ (right).
Right panel: calculation with Model 2 at $p_t = 900\, \mbox{MeV}/c$;
results at $p_t = 300\, \mbox{MeV}/c$ are similar to Model 1.
\label{fashbt}}
\end{figure}
One quantifies the azimuthal dependence of HBT radii by 
decomposing them into Fourier series. Up to leading oscillating 
terms this reads
\begin{equation}
\begin{array}{ll}
R_o^2(\phi) = R_{o,0}^2 + 2 R_{o,2}^2 \cos(2\phi) + \dots\, ,  &
\qquad
R_s^2(\phi)  = R_{s,0}^2 + 2 R_{s,2}^2 \cos(2\phi) + \dots\, , \\
R_{os}^2(\phi) = 2 R_{os,2}^2 \sin(2\phi) + \dots \, , & 
\qquad
R_l^2(\phi)  = R_{l,0}^2 + 2 R_{l,2}^2 \cos(2\phi) + \dots \, .  
\end{array}
\label{r2}
\end{equation}
We will be interested in the dependence of second-order 
amplitude of the outward radius $R_{o,2}^2$ and the sideward radius
$R_{s,2}^2$ on the spatial anisotropy $a$ and 
flow anisotropy $\rho_2$. We would like to eliminate their scaling with the 
absolute transverse size of the fireball, so we will divide
them by the respective average terms $R_{i,0}^2$. 
 
This is plotted in Figure~\ref{fashbt}. The good news from this figure 
is that almost in all cases the normalised oscillation amplitude 
$R_{i,2}^2/R_{i,0}^2$ depends mainly on spatial anisotropy and only 
weakly on flow anisotropy. The only found exception is $R_s^2$ in Model 2:
its oscillation is dictated by flow asymmetry at high $p_t$. 
At low $p_t$, the homogeneity regions are mainly characterised by 
the shape of the fireball. At higher $p_t$, they become more
sensitive to the flow. In case of Model 2 this can even lead to a change of
sign of the amplitude $R_{s,2}^2$ \cite{prpaper}.

\subsection{Comparison with data}

\begin{figure}[h]
\begin{minipage}[b]{10cm}
\centerline{\includegraphics[width=9.9cm]{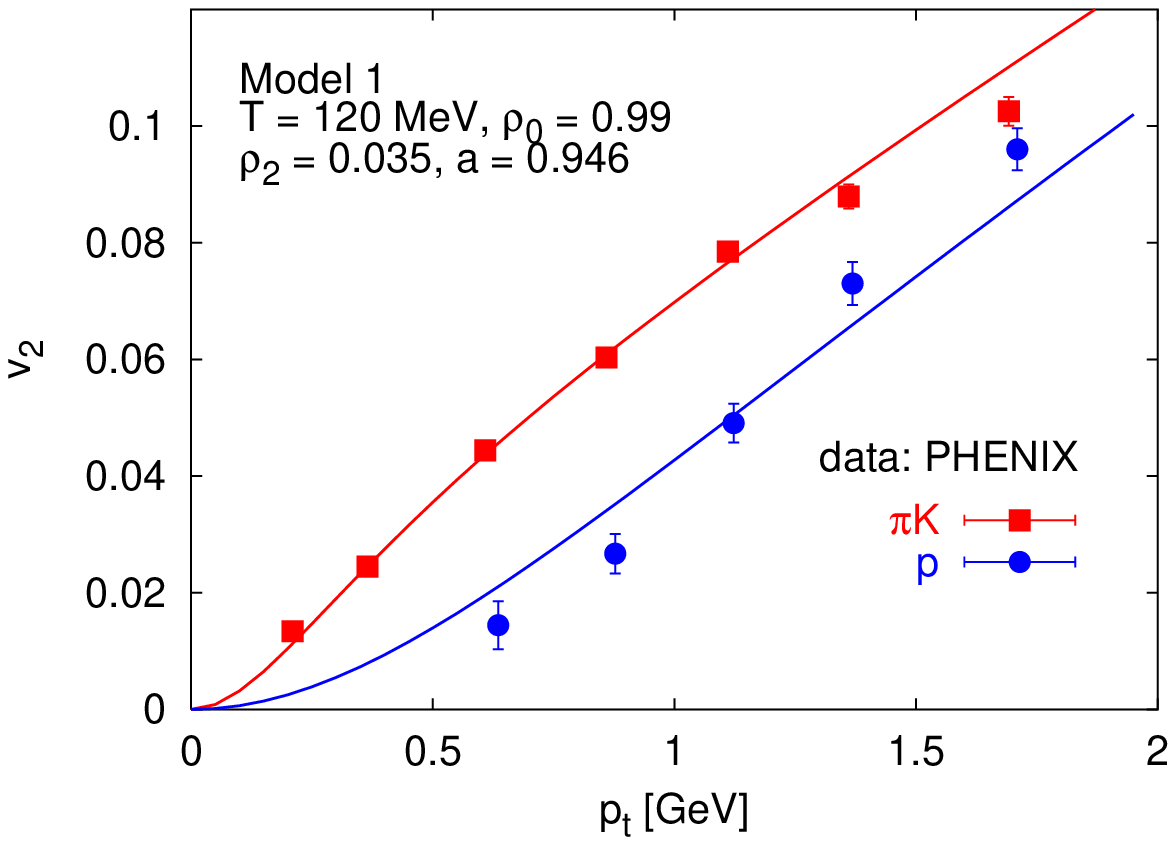}}
\end{minipage}
\begin{minipage}[b]{5.9cm}
\caption{%
Elliptic flow coefficient $v_2$ measured by the PHENIX collaboration
in Au+Au collisions
at $\sqrt{s} = 200~A\mbox{GeV}$ for the centrality class 20--30\%
\cite{phv2} as a function of $p_t$ for a pions and kaons (squares) and
for identified protons (circles). Theoretical curves are obtained with the 
Model 1 using parameter values indicated in the figure.
\label{v2fit}}
\end{minipage}
\end{figure}
\begin{figure}[h]
\centerline{\includegraphics[width=13cm]{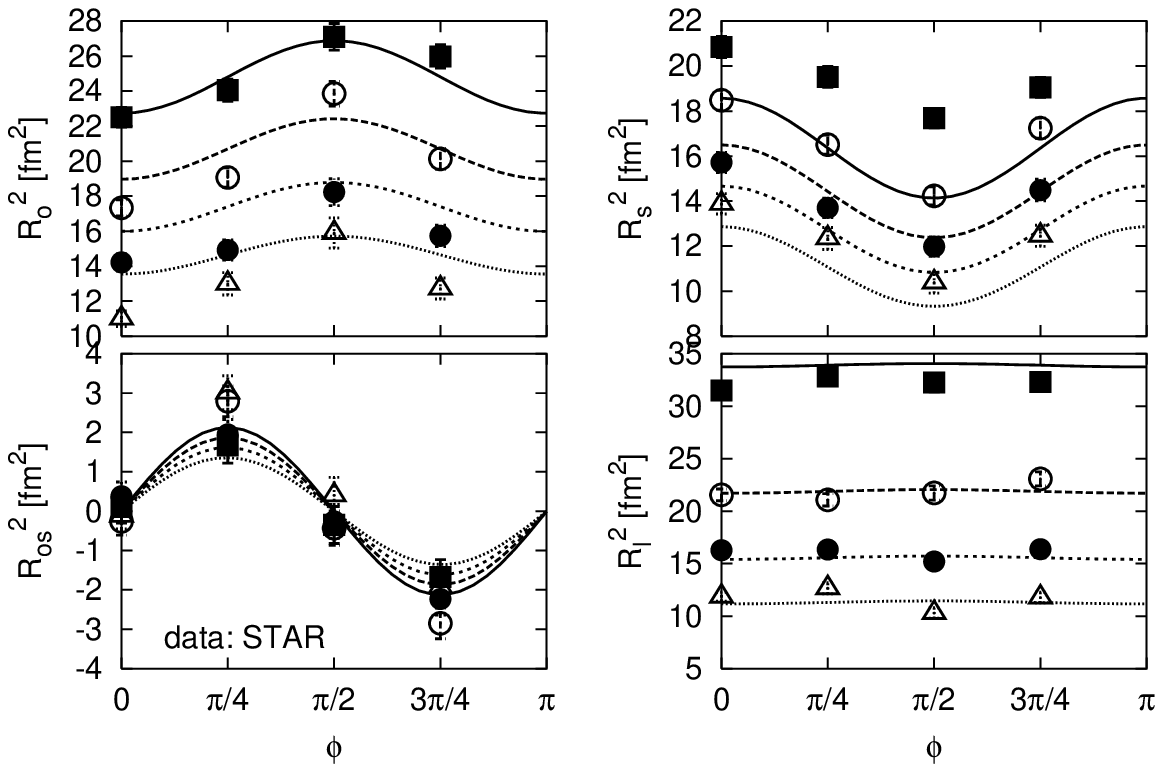}}
\vspace*{-1cm}
\caption{%
Azimuthal dependence of the HBT radii at midrapidity from Au+Au collisions
at 200 $A$GeV and centrality class 20--30\%
measured by the STAR Collaboration \cite{stahbt}. 
From highest till lowest the data are taken for $K_t$ values: 0.2, 0.3, 0.4, 
and 0.52 GeV/$c$. Theoretical curves
are calculated with Model 1 with the parameter values: $T = 120\, \mbox{MeV}$,
$\rho_0 = 0.99$, $\rho_2 = 0.035$, $a = 0.94646$, $R=9.4\, \mbox{fm}$,
$\tau_0 = 5.0\, \mbox{fm}/c$, $\Delta\tau = 2.9\, \mbox{fm}/c$.
\label{ashbt}}
\end{figure}
By comparing to data one can try to distinguish whether we observe an 
in-plane or out-of-plane elongated fireball and which of the two 
introduced models
is better. In Figure~\ref{v2fit} I show $v_2(p_t)$ compared to 
curves calculated with Model 1. Within this model we obtain an 
out-of-plane elongated source ($a<1$), i.e., a source which remembers
its original deformation. Recall, however, that the same theoretical curves
can be obtained in Model 2 if we change $a$ into $1/a$, so we would have 
an in-plane elongated source. 

The key to resolve this ambiguity is a fit to the data on azimuthal
dependence of  HBT radii. I can fit their oscillation well with 
Model 1 if I assume parameters from the fit to $v_2$; this is shown
in Figure~\ref{ashbt}. I checked that in Model 2 the amplitude of 
oscillations is opposite to what the data show \cite{prpaper}. 
The conclusion is
that the observed fireball is {\em out-of-plane} elongated and 
that Model 1 reproduces data better than Model 2.

With the used ``conventional'' values of temperature and average
radial flow gradient $\rho_0$ I could not reproduce the absolute size 
and $K_t$ dependence of the sideward HBT radius, but note that 
this is the same kind of problems as was observed in central collisions.


\section{CONCLUSIONS}

We do see a strong collective transverse expansion at RHIC (and 
at SPS) because the transverse size of the source is much larger than
the original nuclei. The total lifetime seems to be short: this is 
indicated by the size of $R_l$ when interpreted in Bjorken scenario
and by the fact that in non-central collisions we observe an 
out-of-plane elongated source so there wasn't enough time to
wash away this anisotropy. On the other hand, the lifetime must also 
be long enough to allow the fireball for a massive transverse
expansion. This, together with the failure of the blast-wave model 
in fitting the HBT radii, indicates that the model does not 
provide a satisfactory description of the decoupling of particles. 
As the freeze-out temperature is always obtained in a specific model 
there is no conclusion about its particular value.

\section*{Acknowledgements}
I thank the organisers for the invitation and for the stimulating 
environment created at the conference. I am grateful to Scott Pratt 
and Evgeni Kolomeitsev for inspiring discussions and to Mike Lisa and
Fabrice Reti\`ere for critical reading and comments to the manuscript. 
This research was supported by a Marie Curie Intra-European Fellowship
within the 6th European Community Framework Programme.

\end{document}